\documentclass[preprint,showpacs,pre,amsmath,amssymb,endfloats*]{revtex4}
\usepackage{graphicx}
\usepackage{dcolumn}
\begin{document}

\preprint{Frustrated colloidal glass}

\title{Induced glassy behavior in the melt of glycerol and aerosil dispersions}

\author{Dipti Sharma and Germano S. Iannacchione}
\affiliation{Department of Physics, Worcester Polytechnic
Institute, Worcester, Massachusetts 01609, USA}

\date{\today}


\begin{abstract}
A high-resolution calorimetric spectroscopy study has been
performed on pure glycerol and colloidal dispersions of an aerosil
in glycerol covering a wide range of temperatures from $300$~K to
$380$~K, deep in the liquid phase of glycerol. The colloidal
glycerol+aerosil samples with $0.05$, $0.10$, and $0.20$ mass
fraction of aerosil reveal glassy, activated dynamics at
temperatures well above the $T_g$ of the pure glycerol. The onset
of glass-like behavior appears to be due to the structural
frustration imposed by the silica gel on the glycerol liquid. The
aerosil gel increases the net viscosity of the mixture, placing
the sample effectively at a lower temperature thus inducing a
glassy state. Given the onset of this behavior at relatively low
aerosil density (large mean-void length compared to the size of a
glycerol molecule), this induced glassy behavior is likely due to
a collective mode of glycerol molecules. The study of frustrated
glass-forming systems may be a unique avenue for illuminating the
physics of glasses.
\end{abstract}



\maketitle


\section{INTRODUCTION}
\label{sec:intro}

The glass-forming liquids are interesting materials for study and
have continually drawn the attention of researchers because of
their unusual material properties
\cite{Simon97,Richert98,Birge97,Weyer97,Alig97}. These materials
show different thermal transitional and dynamical effects when
they enter the glassy state. That is, the glassy state is governed
by the disorder of the molecules when quenched rapidly below the
glass transition temperature ($T_g$). Much attention has been
devoted to the study of pure glass-forming liquids such as
glycerol \cite{Richert98,Prevosto02,Sharma03} since the glassy
state of the pure liquid can be achieved with ease just by cooling
below $T_g (\approx 190$~K for glycerol). It remains unclear
whether the glass state is due to molecular interactions or a
collective phenomena of many molecules that prevents
crystallization and "freezes" the sample into an amorphous state.
One avenue to illuminate this phenomena has been the study of
mixtures of glass-forming liquids or other solvents to isolate any
specific molecular interactions. However, only a few such studies
have been performed to date, and these have not settled the
central question \cite{Yu04}.

An extension of the idea of using mixtures would be to employ a
colloidal dispersion in a glass-forming liquid to introduce
structural frustration. Such a gel system would introduce
frustration in a controlled way by simply increasing the density
of the colloidal particles. This would have the advantage of
increasing the effective viscosity of the mixture and would have
an associated mean void length. The perturbations observed for a
given mean void length could then be compared to a typical
molecule size of the glass-forming liquid to determine the extent
of any collective mode. The colloidal gel could be obtained using
aerosil particles, a technique that has already found use in the
study of quenched random disorder at liquid crystal phase
transitions and has been shown to involve coupling of the aerosil
to the dynamics of the host fluid \cite{Germano98,Germano03}.

The most common technique for studying different transitions is
differential scanning calorimetry (DSC) and modulation
differential scanning calorimetry (MDSC) techniques, which have
been widely used in the study of glasses \cite{Donth97,Sharma00}.
However, the necessary fast scan rates severely distort the
behavior of slowly relaxing systems. To measure the heat capacity
directly and over a wide range of frequencies under near
equilibrium conditions, ac or modulation calorimetry (MC) has
proven to be of great benefit for thermodynamic studies
\cite{Simon97,Germano97b,Kraftmakher02,Gmelin97}. Our interest is
to use a MC technique on frustrated glass-forming liquid systems
to probe energy relaxations via calorimetric spectroscopy.

This paper presents the results of a calorimetric spectroscopy
study of colloidal dispersions of aerosil particles in glycerol as
a function of aerosil density. The addition of aerosil introduces
glass like dynamics with relaxations on the order of a second
beginning at void length scales on the order of $100$~nm. This
indicates that the induced glassy behavior is due to a collective
mode of many glycerol molecules. A theoretical review of
calorimetric spectroscopy and the relevant glass dynamics is
presented in Sec.~\ref{sec:theory}, and the experimental details
are described in Sec.~\ref{sec:exp}. The results for temperature
and frequency scans as well as an Arrhenius analysis are shown in
Sec.~\ref{sec:results}, with discussion and conclusions drawn in
Sec.~\ref{sec:disc}.


\section{Theoretical Description}
\label{sec:theory}

\subsection{Complex Heat Capacity}

Since the source of any sample's heat capacity are the
fluctuations of the sample's energy, it is natural to consider $C$
as a dynamic quantity. However, for most materials the
characteristic relaxation time of the energy fluctuations are too
short to be sensed by traditional calorimetric techniques.
Nevertheless, a complex heat capacity may be defined containing a
real ($C'$) and imaginary ($C''$) component indicating the storage
(capacitance) and loss (dispersion) of the energy in the sample in
a direct analog to a complex permittivity. Using a linear response
approach, the relaxing enthalpy fluctuations (at constant
pressure) denoted as $\delta H_R$ defines an enthalpy correlation
function as $\langle \delta H_R(0) \delta H_R(t) \rangle$. The
complex heat capacity at a given frequency $\omega$ is then given
by
\begin{equation}
\label{eq:Cp}
   C_p(\omega) =
   C_p'(\omega) - i \int_0^\infty \omega k_B \beta^2 \langle\delta H_R (0)
   \delta H_R (t)\rangle dt
\end{equation}
where $\beta = (k_B T)^{-1}$ and the time integral multiplying the
imaginary number is $C_p^{''}$. For energy dynamics characterized
by a single mode of frequency $\omega_m$, the real part has two
asymptotic limits, fast ($\omega \rightarrow \infty$ or $\gg
\omega_m$, denoted by $C_p^\infty$) and static ($\omega
\rightarrow 0$ or $\ll \omega_m$, denoted by $C_p^0 = C_p^\infty +
k_B\beta^2\langle \delta H_R^2 \rangle$) compared to this mode.
The imaginary part would exhibit a peak at $\omega_m$ that would
be the inflection point of the smooth roll over between the two
frequency limits of $C_p'$.

\subsection{Modulation Calorimetry}

Modulation calorimetry allows one to make frequency dependent
$C_p$ measurements and so access calorimetric spectroscopy. In
modulation calorimetry, a small oscillating heating power $P(t) =
P_0\exp(i\omega t)$ is applied to the sample+cell and induces an
\textit{rms} temperature rise as well as small temperature
oscillations. Note, in what follows the cell is considered to
consist of the actual sample holder and the attached heater and
thermometer. For a sample of finite thermal conductivity and a
sample+cell experimental arrangement of a thickness less than the
thermal diffusion length, the heat flow equations may be set up
considering the sample+cell as a single thermal mass (i.e.,
geometry independent) \cite{Sullivan68}. The temperature
oscillation amplitude $T_{ac}$ is given to second order by
\begin{equation}
\label{eq:Tac}
   T_{ac} = \frac{P_0}{\omega C} \left(1 + ({\omega \tau_e})^{-2} + \omega^2 \tau^2_{ii}
   + \frac{2R_s}{3R_e} \right)^{-1/2} ,
\end{equation}
where $P_0$ is the amplitude of the applied heating power,
$\omega$ is the angular frequency, $C = C_s + C_c$ is the total
heat capacity of the sample and cell, and $T_{ac}$ is the
amplitude of the temperature oscillations. In addition, there are
two important thermal relaxation times, the external relaxation
time constant $\tau_e = R_eC$ and the internal relaxation time
constant $\tau_{ii}^2 = \tau_s^2 +\tau_c^2$, which is the sum of
square internal time constants of the sample and cell ($\tau_s =
R_sC_s$ and $\tau_c = R_cC_c$). Here, $R_s$ is sample's thermal
resistance and $R_e$ is the external thermal resistance to the
bath. There is also a phase shift between the applied heat and
resulting temperature oscillations $\Phi$ but it is more
convenient to define a reduced phase shift $\phi = \Phi + \pi/2$
since for frequencies below $1/\tau_{ii}$, $\Phi \approx -\pi/2$.
The reduced phase shift, to the same accuracy as
Eq.~\eqref{eq:Tac}, is given by
\begin{equation}
  \label{eq:Tanphi}
  \tan(\phi) = (\omega\tau_e)^{-1} - \omega\tau_{i}
\end{equation}
where $\tau_i = \tau_s + \tau_c$. The different internal
relaxation times that enter in Eq.~\eqref{eq:Tac} and
\eqref{eq:Tanphi} are related by $\tau_{ii}^2 = \tau_i^2 +
2\tau_s\tau_c$ and must be taken into account in order to extract
the complex heat capacity.

Noting that $1 + \tan^2(\phi) = \cos^{-2}(\phi)$,
Eq.~\eqref{eq:Tanphi} can be substituted into Eq.~\eqref{eq:Tac}
to give
\begin{equation}
\label{eq:C}
   C =
   C^\ast \left[ \frac {1}{\cos^2(\phi)} - 2{\tau_c}{\tau_s}{\omega}^2
   + 2 \frac{\tau_i}{\tau_e} + \frac{2R_s}{3R_e} \right]^{-1/2} ,
\end{equation}
where $C^\ast \equiv P_0 / (\omega T_{ac})$ is defined for ease of
notation. Factoring out a $\cos(\phi)$ gives
\begin{equation}
\label{eq:ReC1}
   C' = C^\ast \cos(\phi) f(\omega)
\end{equation}
where the total heat capacity is identified as the real part of
the complex heat capacity. The function $f(\omega)$ is then given
by
\begin{equation}
\label{eq:fw}
   f(\omega) = \left[ 1 + 2\left( \frac{\tau_i}{\tau_e}
   + \frac{R_s}{3R_e} - {\tau_c}{\tau_s}{\omega}^2 \right)\cos^2(\phi)
   \right]^{-1/2},
\end{equation}
and may be regarded as a correction for a sample and cell having
comparable finite thermal relaxation times.

For most materials away from any phase transition, the imaginary
part of the heat capacity is essentially zero at typical heating
frequencies ($\leq 15$~mHz). This quantity may be derived from
Eq.~\eqref{eq:Tanphi} by substituting the definitions of the
relaxation times and Eq.~\eqref{eq:ReC1} for the heat capacity.
This gives
\begin{equation}
\label{eq:ImC1}
   C^{''} =
   C^\ast \sin(\phi) g(\omega) - (\omega R_e)^{-1} \approx 0.
\end{equation}
where a frequency dependent correction function is introduced as
\begin{equation}
\label{eq:gw}
   g(\omega) = f(\omega)\left( 1 + \frac{\omega \tau_i}{\tan(\phi)} \right)
\end{equation}
that also accounts for the comparable relaxation times of the
sample and cell.

Typically, the cell's geometry and mass may be controlled and so
its thermal relaxation time can be made much less than that of the
sample, $\tau_c \ll \tau_s$. In this case, the two internal
relaxation times are approximately equal. Taking $\tau_{ii} =
\tau_i$, the correction function $g(\omega)$ does not change but
the $\omega^2$ term vanishes in $f(\omega)$. The frequency
dependence for both correction functions enter through the reduced
phase shift. At sufficiently low heating frequencies $\omega <
\tau_e^{-1}$, $\phi$ approaches $\pi/2$ ($\Phi$ approaches $0$)
and $f(\omega) \simeq g(\omega) \approx 1$, which reduces the real
and imaginary heat capacities in Eqs.~\eqref{eq:ReC1} and
\eqref{eq:ImC1} to
\begin{equation}
\label{eq:ReC2}
   C' = C^\ast \cos(\phi)
\end{equation}
and
\begin{equation}
\label{eq:ImC2}
   C^{''} = C^\ast \sin(\phi) - (\omega R_e)^{-1} .
\end{equation}
where any energy dynamics or dispersion in the sample will be
indicated by a non-zero value of $C^{''}$.

\subsection{Brief Review of Glass Dynamics}

Glass-forming or super-cooled liquids exhibit dynamics due to
inhibited structural relaxations that diverge upon approaching the
glass-transition until the system falls from ergodicity.
Glass-formers are characterized by a rapid increase in viscosity
(from $10^{-2}$ poise above to $10^{15}$ poise below the glass
transition temperature) with decreasing temperature, which
reflects the rapidly shifting timescale of relaxation dynamics of
the liquid. The slow molecular motion, known as
$\alpha$-relaxation, has been well characterized on timescales
between $100$~ps and longer \cite{Yang95,Glorieux02} and this
stops completely below the glass transition temperature. The fast
motion ($\beta$-relaxation) is predicted to exist at timescales
between $1$ and $100$~ps \cite{Franosch98}, and persists even in
the glassy state. However, $\beta$-relaxation is difficult to
determine unambiguously by experiment and its physical meaning is
debated. Qualitatively, $\alpha$-relaxation can be thought of as
the collective motion of many particles while $\beta$-relaxation
is the motion of a single molecule rattling within a cage of
nearest neighbors. Fundamentally, the issue of glasses centers on
whether the glassy behavior is due to "stuck states" of a
collection of molecules or due to molecular frustration preventing
the crystallization of the material.

Quantitatively, glassy dynamics typically obey
Vogel-Fulcher-Tammann (or Arrhenius) behavior indicating that the
dynamics are energetically activated \cite{Vogel21,Fulcher26}.
According to Arrhenius behavior, the relaxation time of glass
forming liquids can be given by
\begin{equation}
\label{eq:tau}
   \tau = \tau_o \exp( \beta \Delta E )
\end{equation}
where $\tau$ is the relaxation time of some structural
fluctuation, $\tau_0$ is the high-temperature limit of this
fluctuation, $\Delta E$ is the activation energy, and $\beta =
(k_BT)^{-1}$ as usual. This activated relaxation mode can become
very long and complicate the interpretation of thermodynamic
measurements. Glycerol has been a particularly well studied
example of such glass-forming liquids that exhibit all of the
above characteristics \cite{Simon97,Richert98,Yu04}.


\section{Experimental}
\label{sec:exp}

\begin{figure}
\includegraphics[scale=0.3]{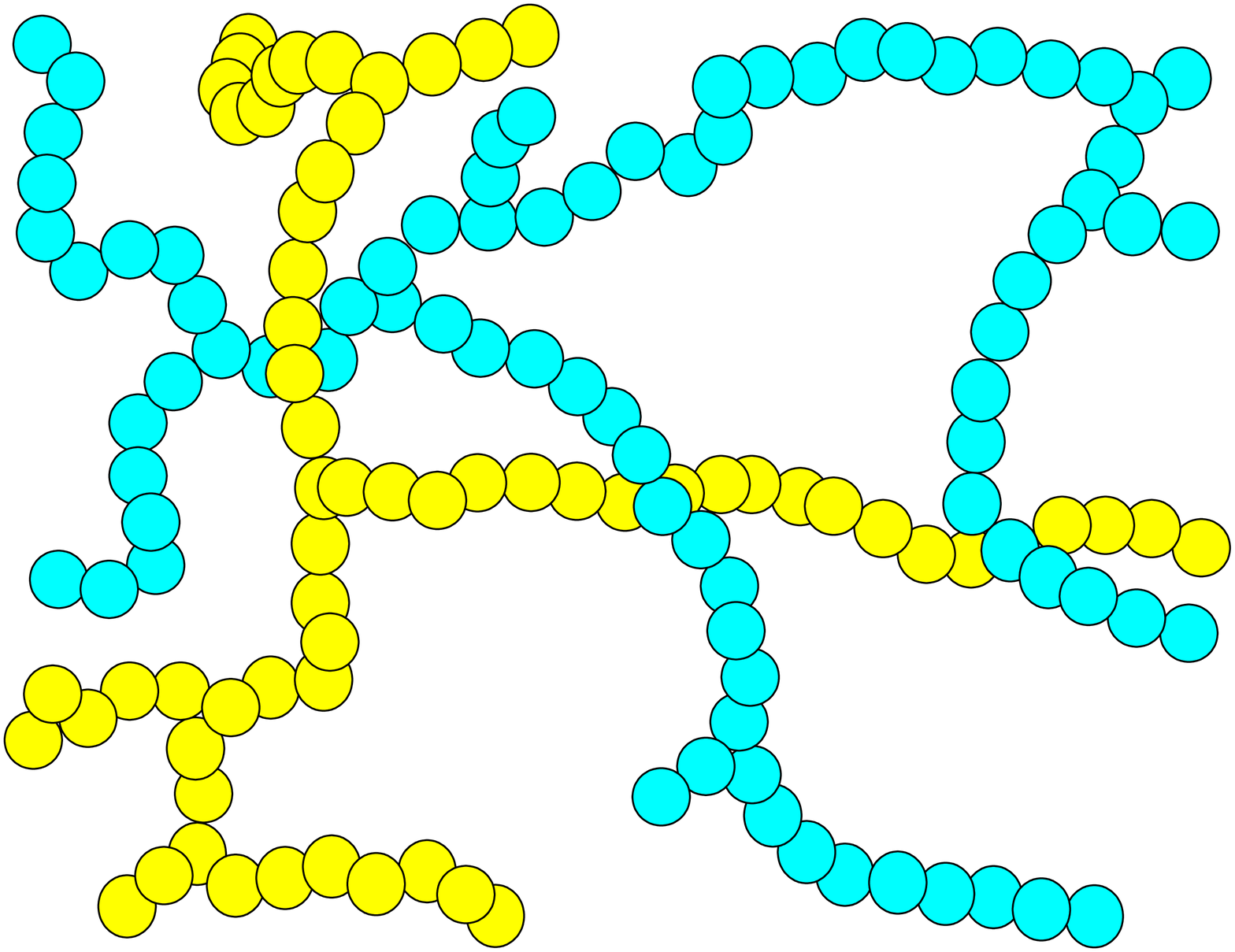}
\caption{ \label{SilCartoon} Cartoon depicting the fractal gel
formed by the diffusion-limited aggregation (hydrogen-bonding) of
aerosil nano-particles in an organic solvent. However, in glycerol
H-bonding between the glycerol and aerosil is also possible and
may change the "pearl-necklace" nature of the aerosil dispersion.
In any case, the distribution of aerosil should remain relatively
random. }
\end{figure}

The pure glycerol obtained from Aldrich was used after carefully
degassing at $323$~K for $\gtrsim 2$~hours because of its
hygroscopic nature. The pure glycerol has a molecular weight of
$M_w = 92.09$~g~mol$^{-1}$, a density of $\rho_g =
1.26$~g~cm$^{-3}$, and a nominal glass transition temperature of
$T_g \simeq 195$~K. The hydrophilic type-300 aerosil silica nano
particles obtained from Degussa \cite{Degussa} were thoroughly
dried at $\sim 573$~K under vacuum for $\sim 2$ hours prior to
use. The specific surface area of the aerosils measured by the
manufacturer via BET nitrogen isotherms is $a =
300$~m$^2$~g$^{-1}$ and each aerosil sphere is roughly $7$~nm in
diameter. However, SAXS studies have shown that the basic aerosil
unit consists of a few of these spheres fused together during the
manufacturing process \cite{Germano98}. Typically, the hydrophilic
nature of the aerosils arises from the hydroxyl groups covering
the surface and allows the aerosil particles to hydrogen bond to
each other. This type of bonding is weak and can be broken and
reformed, which leads to the thixotropic nature of gels formed by
aerosils. The aerosil gelation in an organic solvent occurs via a
diffusion-limited aggregation process resulting in a fractal gel
of $d_f \simeq 2.4$ \cite{Roshi05}. See Fig.~\ref{SilCartoon}.
However, pure glycerol is a hydrogen-bonding liquid that may alter
the dispersion of aerosil from that shown in
Fig.~\ref{SilCartoon}. Each glycerol+sil colloidal dispersion
sample was created by mixing appropriate quantities of aerosil and
glycerol together with spectroscopic grade (low-water content)
acetone that was subsequently evaporated slowly away. The
resulting mixtures were then annealed and degassed at $323$~K for
$1$~hour. The resulting samples appear by visual inspection to be
very similar to those made in an organic solvent like a liquid
crystal. In addition to the pure glycerol sample, three dispersion
samples were prepared with a mass fraction of aerosil of $0.05$,
$0.10$, and $0.20$. In addition, these samples may be
characterized by the conjugate density $\rho_S$ defined as the
mass of aerosil per open (glycerol) volume that allows one to
determine the mean-void length $l_0 = 2 / a
\rho_S$~\cite{Germano03}. The characterization of the aerosil by
$\rho_S$ and $l_0$ does not depend on the details of the
dispersion structure, only that it be random. It is likely that
the aerosil dispersion remains random in glycerol. See
Table~\ref{tab:sum} for a summary of these parameters.

High-resolution ac calorimetry was performed using a home-built
calorimeter at WPI. The sample cell consisted of a silver
crimp-sealed cup+lid where cup dimensions are $\sim 12$~mm
diameter and $\sim 0.5$~mm thick and lid dimensions are $\sim
12$~mm diameter and $\sim 0.1$~mm thick (closely matching the
dimensions of the heater). The average mass of the silver cell was
$0.135$~g and the sample (glycerol+aerosil) was $36$~mg and the
total mass of cell+sample did not deviate by more than $\sim
35$~mg between the different samples. After the sample was
introduced into a cell and the lid crimp sealed to the cup under
pressure a $120$-$\Omega$ strain-gauge heater and $1$-M$\Omega$
carbon-flake thermistor were attached. The cell was then mounted
in the calorimeter, the details of which have been described
elsewhere \cite{Yao98}. In the ac-mode, power is input to the cell
as $P_0 e^{i\omega t}$ resulting in temperature oscillations with
amplitude $T_{ac}$ and a relative phase shift of $\phi$. From
these parameters and the all necessary finite conductivity
corrections, the real and imaginary heat capacity were determined.
All cells studied in this work closely matched each other in
dimension and mass to better than 5\% in order to better isolate
effects introduced by the aerosil.

Three experiments were performed: temperature, time, and frequency
scans with the following protocols. For temperature scans,
measurements were made on a freshly mounted sample at a constant
heating frequency of $15$~mHz as the cell first heated then
immediately cooled between $300$ and $380$~K using a constant rate
of $\pm 2$~K~hr$^{-1}$. After the cell had cooled back to the
initial temperature of $300$~K, a time scan was preformed to
monitor the heat capacity as it relaxed back to the initial value.
Frequency scans were then performed at fixed temperatures from
$300$ to $380$~K in steps of $20$~K over the frequency range from
$1$ to $2000$~mHz (i.e., $\omega$ from $0.0063$ to
$12.6$~s$^{-1}$).


\section{RESULTS}
\label{sec:results}

\subsection{Temperature and time scans}

Heating and cooling scans for all samples are shown in the
Fig.~\ref{CPvsT}. From the $C_p$ data, clear indications of
hysteresis are observed for the pure glycerol in that the values
of $C_p$ on cooling do not reproduce the values observed on the
heating scan of a freshly loaded sample. For the glycerol+sil
samples, this hysteresis decreases with increasing aerosil
content. This effect may be quantified by the area of this
hysteresis loop ($\Delta H_{hyst}$) that decreases in a near
linear fashion with the mass fraction of aerosil. This behavior is
in marked contrast to that of the reduced phase shift. For pure
glycerol $\phi$ reproduces itself between heating and cooling, but
$\phi$ begins to take on increasing hysteresis like that described
for $C_p$ but with \textit{increasing} aerosil content.

The measurements made as a function of time, not shown here, upon
returning to the starting temperature clearly indicate a very
slow, nearly linear, relaxation back to the initial $C_p$ values.
For the pure glycerol, this was observed to take $2$ to $3$ days;
while for the $0.10$ dispersion sample, $5$ to $6$ days were
required for the sample to recover. This time increased with
increasing aerosil content. Clearly, long relaxations persist at
these high temperatures for glycerol and the character of this
relaxation changes with the addition of the aerosil gel. These
extremely slow relaxations (on the order of days) are of unclear
origin.

\subsection{Frequency scans}

\begin{figure}
\includegraphics[scale=0.45]{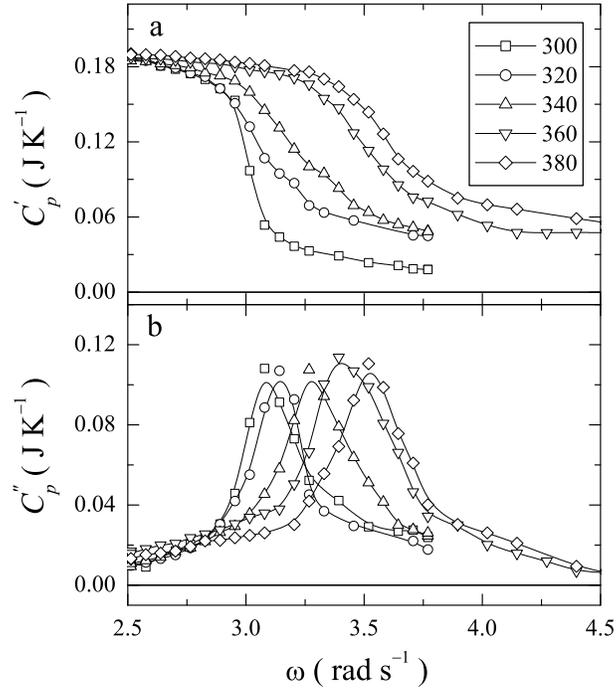}
\caption{ \label{CPvsOMEGA} Real (a) and imaginary (b) heat
capacity as a function of heating frequency for the $0.10$ sample
upon heating a fresh sample to selected temperatures from $300$ to
$380$~K. The lines are guides to the eye and inset denotes the
different temperatures in kelvin. }
\end{figure}

To probe the slow dynamics of the pure and dispersion samples,
frequency scans were performed from $2000$~mHz down to $1$~mHz.
The pure glycerol frequency scans exhibited a dispersion peak in
the imaginary heat capacity at a frequency ($\omega_p$) coincident
with the inflection point of the real heat capacity roll-off. This
dispersion peak remained stationary at $\omega_p \simeq
0.5$~s$^{-1}$ as the temperature progressively increased from
$300$~K to $380$~K. In contrast, the dispersion peak for the
glycerol+sil samples clearly indicate a shifting dispersion peak
with increasing temperature. An example of the resulting real and
imaginary heat capacities after all calibrations and internal
corrections were made is shown in Fig.~\ref{CPvsOMEGA} for the
$0.10$ sample at five temperatures. The imaginary heat capacity
shows a dispersion peak at a frequency that again coincides with
the inflection point of the real heat capacity roll-off. However,
as the temperature increases progressively to $380$~K, the
dispersion peak shifts towards higher frequencies consistent with
results observed in the glassy state of pure glycerol
\cite{Simon97}.

\begin{figure}
\includegraphics[scale=0.45]{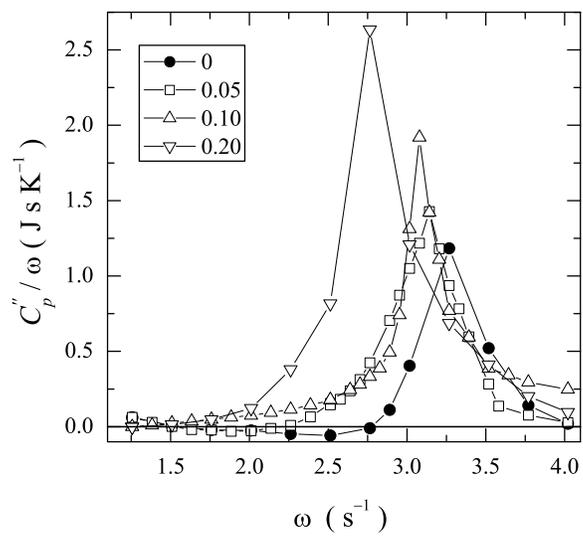}
\caption{ \label{IMCvsOMEGA} Dispersion peak scaled by the
frequency for all samples at $300$~K. The integration of this peak
yields the total dispersion or loss heat capacity, $\Delta
C_p^{''}$. The inset lists the mass fraction of aerosil for each
sample. }
\end{figure}

The evolution of the $C_p^{''}/\omega$ peak at $300$~K as a
function of frequency is given in Fig.~\ref{IMCvsOMEGA} for
several aerosil concentrations. The dispersion peak increases in
height and shifts to lower frequencies with increasing amounts of
aerosil. The integration of $C_p^{''}/\omega$ over the range of
frequency covered yields the total dispersion heat capacity of
this relaxation mode ($\Delta C_p^{''}$), which linearly increases
with aerosil concentration.

\begin{figure}
\includegraphics[scale=0.45]{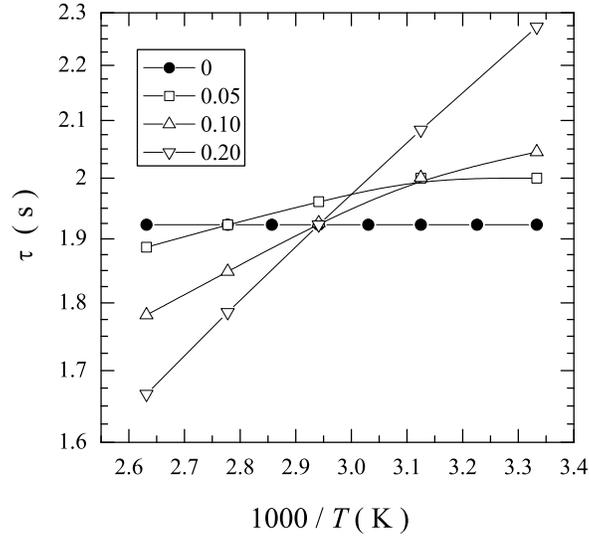}
\caption{ \label{TAUvsT} Semi-log plot of relaxation time $\tau =
1/\omega_p$ (taken from the frequency of the peak in $C_p^{''}$)
versus inverse temperature for each sample studied. The inset
lists the mass fraction of aerosil. The slope provides the
activation energy $\Delta E$ and the intercept at $1000/T = 0$
provides the high-temperature relaxation time $\tau_0$ of the
samples. }
\end{figure}

The frequency of the imaginary heat capacity dispersion peak
$\omega_p$ determines a characteristic relaxation time for this
dynamic mode as $\tau_r = 1/\omega_p$. To establish whether these
relaxations are energetically activated, a semi-log plot of
$\tau_r$ against $1000/T$ was made, which would reveal a linear
region whose slope is directly related to the activation energy
$\Delta E$. See Eq.~\eqref{eq:tau}. Such a plot is shown in
Fig.~\ref{TAUvsT}. The pure glycerol does not exhibit activated
dynamics and the relaxations for the glycerol+sil dispersions
become progressively activated at progressively lower temperatures
representing the onset of Vogel-Fulcher or Arrhenius behavior. It
is clear that as the aerosil content increases, the activation
energy (high-temperature relaxation time) of the glycerol+sil
system continuously increases (decreases). Table~\ref{tab:sum}
summarizes the results of the dispersion peak integration as well
as the Arrhenius analysis.


\section{DISCUSSION AND CONCLUSIONS}
\label{sec:disc}

\begin{figure}
\includegraphics[scale=0.45]{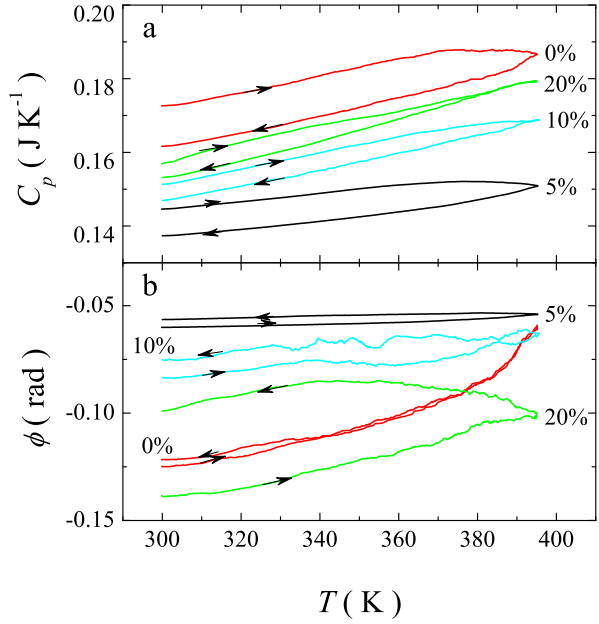}
\caption{ \label{CPvsT} Heat capacity (a) and phase shift (b)
observed in temperature scans for glycerol+sil mixtures containing
$0$, $0.05$, $0.10$, and $0.20$~\% aerosil fraction by mass.
Arrows indicate the direction of the scan, all at a rate of
$2$~K~hr$^{-1}$, with the cooling scan immediately following the
heating. All scans were performed using a heating frequency of
$15$~mHz. Note the hysteresis in $C_p$ that progressively
decreases with increasing silica content. }
\end{figure}

Since the sample and cell configuration was constructed to be
nearly identical for all the samples studied, the differences
observed with the introduction of aerosil is particularly
significant. This is especially true since the thermal
conductivity of the silica is greater than the glycerol and its
increasing content should have improved the internal thermal
conductivity of the sample. Also, since both the aerosil and
glycerol can hydrogen-bond to themselves and each other, the
aerosil surface is likely coated by a bound layer of glycerol.
This would strongly couple the two components of this dispersion.
At room temperature and above, pure glycerol behaves as an
isotropic liquid but still retains some strong molecular
interactions to induce the hysteresis shown in Fig.~\ref{CPvsT}.
The mass of the cell+sample was checked repeated at various stages
during the scan protocol and there were no significant changes.
Thus, desorbing and absorbing moisture during the thermal scans
cannot account for the observed hysteresis behavior. However, this
hysteresis relaxes very slowly, on the order of days, and so it
does not seem to be connected to the dynamics observed by the
frequency scans (on the order of a second) for the pure glycerol
or the glycerol+sil samples.

\begin{figure}
\includegraphics[scale=0.45]{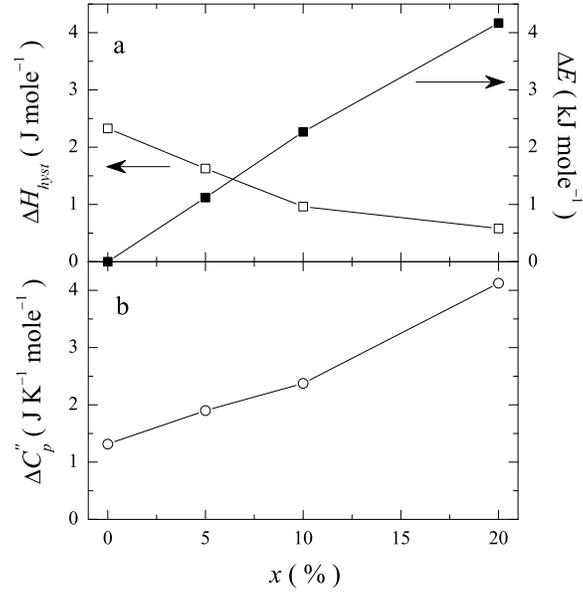}
\caption{ \label{DEvsRHO} Density dependence of the energy
dynamics. Panel (a) shows the hysteresis energy ($\Delta
H_{hyst}$) on left and activation energy ($\Delta E$) on right,
and panel (b) depicts the total dispersion heat capacity ($\Delta
C_p^{''}$) of samples containing $0$, $5$, $10$, and $20$\% silica
by mass. }
\end{figure}

As the silica density increases, the effective viscosity of the
colloidal mixture increases and there is also an increase in the
activation energy. Since these experiments were conducted well
above the pure glycerol glass transition temperature of $195$~K,
the induced glassy behavior is likely due to the aerosil-imposed
frustration on a large collection of glycerol molecules. This is
evident considering that the onset of this effect begins even for
the $0.05$ sample, which has a mean void length scale of $\sim
100$~nm. The estimated size of a glycerol molecule from the
specific volume is $\sim 1$~nm. A summary of the thermodynamic
results are given in Fig.~\ref{DEvsRHO} and Table~\ref{tab:sum}.
Clearly there is a strong correlation of the activation energy
$\Delta E$ and the total heat capacity loss $\Delta C_p^{''}$,
both increasing linearly with aerosil content.

The energy dynamics of glycerol containing a colloidal dispersion
of aerosil silica nano particles has been studied by
high-resolution modulation calorimetry. This has revealed a
temperature and silica density dependent complex heat capacity. It
has been observed that under certain experimental conditions of
temperature and frequency, this system shows interesting frequency
dynamics and displays dispersion peaks that shift towards higher
frequency with increasing temperature. These results are
consistent with those found for the glassy state of glycerol
\cite{Simon97}. As the silica density is increased in the
isotropic glycerol, larger dispersion peaks are found at lower
frequencies indicating an increase in the activation energy of
this relaxation. This phenomena may be explained in terms of the
formation of "stuck states" of glycerol in the voids between
glycerol-coated silica particles also manifesting itself in the
increase in the sample's viscosity. This work highlights the
potential of using frustrated glass-forming liquids via colloidal
gel dispersions as a means of gaining insight into the glass
state. Dielectric spectroscopy studies of such systems as well as
x-ray analysis of the structure of the aerosil dispersion would be
of great interest.


\begin{acknowledgements}
We are grateful to C. W. Garland for useful discussions. This work
was supported by the NSF-CAREER award DMR-0092786.
\end{acknowledgements}


\begin{table*}
\caption{ \label{tab:sum} Summary of the calorimetric results for
the glycerol+aerosil samples. Shown are the silica mass fraction
$x = M_{sil}/M_{Total}$, conjugate density $\rho_S$ (in g of
aerosil per cm$^3$ of $glycerol$), mean void length $l_0$ (in nm),
integrated dispersion peak $\Delta C_p^{''}$ (in
J~K$^{-1}$~mole$^{-1}$),the high temperature the activation energy
$\Delta E$ (in kJ~mole$^{-1}$), and the high-temperature
relaxation time $\tau_0$ (in seconds) obtained by an Arrhenius
analysis. }
\begin{ruledtabular}
\begin{tabular}{@{\extracolsep{20 pt}}llllll}
   $x$ & $\rho_S$ & $l_0$ & $\Delta C_p^{''}$ & $\Delta E$ & $\tau_0$  \\
 \hline
    $0$  &  $0.00$  &  $\infty$  &  $1.32$  &  $0.00$  &  $1.92$  \\
 $0.05$  &  $0.07$  &  $101$     &  $1.90$  &  $1.12$  &  $1.36$  \\
 $0.10$  &  $0.14$  &  $48$      &  $2.37$  &  $2.27$  &  $0.92$  \\
 $0.20$  &  $0.32$  &  $21$      &  $4.13$  &  $4.17$  &  $0.49$  \\
\end{tabular}
\end{ruledtabular}
\end{table*}


\end{document}